\documentclass[preprint,showpacs,preprintnumbers,amsmath,amssymb]{revtex4}
 
\usepackage{graphicx}
\usepackage{dcolumn}
\usepackage{bm}
\usepackage{epsfig}

\newcommand{\ybox}[2]   {       
 \begin{center} 
 \resizebox{!}{#1\textheight}   
{\includegraphics{#2.eps}}      
 \end{center}           }

\begin{document}

\preprint{}

\title{Collective behavior in nuclear interactions and shower development}

\author{V.~Canoa$^1$, N.~Armesto$^2$, C.~Pajares$^1$
and R.~A.~V\'azquez$^1$}

\affiliation{$^1$ Departamento de F\'\i sica de Part\'\i culas,
Universidade de Santiago de Compostela, E-15706 Santiago
de Compostela, Spain}

\affiliation{$^2$ Department of Physics, CERN, Theory Division, CH-1211
Gen\`eve 23, Switzerland}

\begin{abstract} 
The mechanism of hadronic interactions at very high energies is still
unclear.  Available accelerator data constrain weakly the forward
rapidity region which determines the development of atmospheric
showers. This ignorance is one of the main sources of uncertainty in
the determination of the energy and composition of the primary in
hadron-induced atmospheric showers. In this paper we examine the
effect on the shower development of two kinds of collective effects in
high-energy hadronic interactions which modify the production of
secondary particles.  The first mechanism, modeled as string fusion,
affects strongly the central rapidity region but only slightly
the forward region and is shown to have very little effect on the
shower development. The second mechanism implies a very strong
stopping; it affects modestly the profile of shower maximum but
broadens considerably the number distribution of muons at ground.
For the latter mechanism, the development of air showers is faster
mimicking a heavier projectile. On the other hand, the number of muons
at ground is lowered, resembling a shower generated by a lighter primary. 
\end{abstract}


\maketitle

\section{Introduction}

The cosmic ray energy spectrum extends up to several $10^{20}$ eV.
The presence of these high energy particles (UHECRs) poses an enigma,
since many good arguments suggest that they should not be observed
because of the GZK cutoff \cite{GZK}.  This apparent contradiction has
stimulated a variety of more exotic explanations of their origin and
composition which is usually known as the GZK puzzle. Both
theoretically and experimentally the problem has been the subject of
considerable efforts \cite{NW}.

One of the major uncertainties is the lack of information about the
composition of the highest energy cosmic rays. Besides being a
difficult, indirect measurement with contradictory experimental
results, it involves extrapolations of hadronic physics models far
beyond the actual measurements. The systematic effects due to the
uncertainty of the hadronic interactions, see e.g. \cite{Luna}, 
at these energies are far from being established.

In Cosmic Ray Physics, it is usually assumed that a nucleus-nucleus
collision occurs as an independent combination of nucleon-nucleon
interactions. This approximation is useful in that it allows to obtain
scaling relations for shower properties like shower maximum and the
muon content, which are roughly obeyed.  But it is not justified at
high energies: a beam nucleon may interact several times with
different target nucleons, which will invalidate the approximation
trivially. A more interesting possibility would be the existence of a
higher degree of collectivity (defined in the sense of a failure in the
independence of the superposition of nucleon-nucleon
interactions). This collectivity should be present in any type of
hadronic collisions, but its enhancement due to the hadron size makes
it more prominent for nucleus-induced reactions. It could eventually
lead to the formation of new nuclear phases during the collision, like
the Quark-Gluon Plasma. The existence or not of such new form of
matter at the SPS at CERN and RHIC at BNL has been widely discussed
\cite{qgpclaims}.  For example, it is of relevance for our study that
some violation of the linear scaling law for the rapidity shift of the
projectile nucleons, which works at lower energies, has already been
observed at RHIC \cite{RHIC}.

Here we will investigate the effects of collective behavior on the
development of high energy cosmic ray showers. Collective behavior
may affect shower development in two different ways. First, it could
lead to changes in the total cross sections. Such changes have been
investigated in some models and found to be negligible, see e.g.
\cite{Nestor}.  Second, it is generally expected that any collective
behavior would affect the spectra of secondary particles; this is the
point that we will try to address in our study, exploring it in two
directions. On the one hand, it may reduce the total multiplicity due
to shadowing. As we will see below this would affect the shower
development only if the multiplicity reduction can be carried over the
whole cascade process \cite{Pajares,Sousa}.  This is, in general, not
expected.  Alternatively, collective effects may affect the secondary
particle spectrum, in particular in the large Feynman-$x$ region,
thereby modifying the inelasticity of the collision. This may have
important consequences in shower development as we will show below.
Other possible consequences of collectivity like heavy flavor
production or modifications of large-$p_t$ particle production are 
expected to have much smaller effect on the shower development.

We will analyze the second possibility within both directions. We will
consider first the possibility of string fusion \cite{fusion,psm} as a
collective effect which strongly reduces the multiplicities in the
small Feynman-$x$ region but modifies only slightly the large
Feynman-$x$ one.  It turns out that this effect produces negligible
modifications in shower development, as expected.  We then consider a
second possibility in which extremely strong collective effects do
modify the stopping power and inelasticity in the
collision. Concretely, it has been suggested recently by Mishustin and
Kapusta \cite{Kapusta} that in the central region of a high-energy
collision, strong chromo-electric fields are formed. They will attract
the forward and backward particles, reducing considerably the rapidity
of the fast--moving, leading secondaries.  This effect is expected to
increase the inelasticity even for peripheral collisions. It will lead
to a small change in the shower depth profile but it will induce a
large broadening of the number distribution of muons at ground.

The plan of the paper is as follows. In the next Section we will
briefly discuss the ingredients present in hadronic models introduced
in the simulation of atmospheric showers. In Section III we will
analyze the influence of the string fusion mechanism on shower
development. The extreme stopping scenario of \cite{Kapusta} will be
discussed in Section IV, and Section V will be devoted to our
conclusions.

\section{Hadronic models}

Most of the models which simulate interactions of nucleons are based
on the Gribov-Regge Theory (GRT) \cite{Gribov}, which most
successfully describes elastic scattering and, via the optical
theorem, the total hadronic cross section as a function of energy. In
GRT the observed rise of the cross section at high energies is a
consequence of the exchange of multiple supercritical Pomerons, the
Pomeron being here the effective exchange. Inelastic processes are
described by cut Pomerons. In the Dual Parton Model (DPM) or
Quark-Gluon String Model (QGSM) \cite{DPM}, each cut Pomeron leads to
two color strings which fragment subsequently into color--neutral
hadrons. The number of exchanged strings (or Pomerons) rises with
energy.  Most models widely used in Cosmic Ray Physics to compute
hadronic collisions at high energies, e.g. \cite{QGSjet,nexus,dpmjet},
are based on these ideas for the low transverse momentum region; for
hard particle production, perturbative QCD is used with a matching
between the soft and hard components.

A possible formulation of GRT can be done \cite{Abramovsky:yw}
assuming that strings have a transverse size $A=\pi r^2$.  As the
number of strings increases with collision energy, mass number of
projectile and target and decreasing impact parameter (increasing
centrality) of the collision, it is expected that the strings overlap
and can no longer be considered to evolve and fragment independently.
The simplest possibility to consider is fusion of strings.  The fusion
probability is determined by some parton or string transverse
dimension and by the corresponding parton-parton cross section.  This
model is the string fusion model (SFM) \cite{fusion} embedded in the
DPM.  In the last version of this model, in the form of the program
PSM \cite{psm}, only fusion of pairs of strings is taken into account
and consecutive fusions of strings are neglected. Therefore, 
the string fusion parameter, the transverse area occupied by
a string, is considered as an effective parameter 
that is taken to be \cite{psm}
$\sigma_{p}=2\pi r^{2}= 7.5 $ mb.  Let us note that for proton-nucleus
collisions the number of strings is much reduced compared to
nucleus-nucleus collisions. Therefore even at the highest cosmic ray
energies the effect of string fusion on proton-air collisions is very
small.

\begin{figure}  
\ybox{0.4}{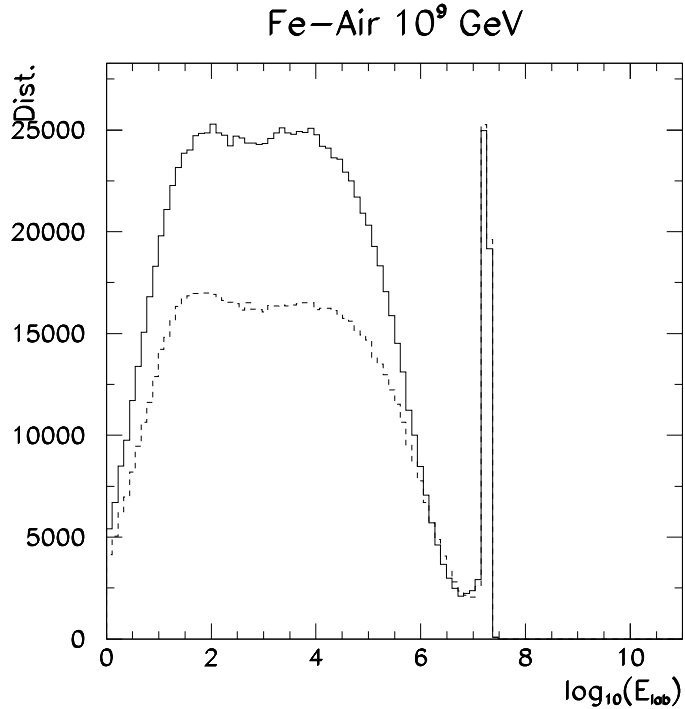}  
\caption{Secondary particle energy distribution for Iron-Nitrogen
  collisions at $10^{9}$ GeV laboratory energy without fusion
  (continuous line) and with fusion (dotted line).}
\label{fig1}
\end{figure}

Consequences of string fusion \cite{fusion,psm} are the reduction of
multiplicities at central rapidities, a slight increase in the
transverse momentum of the produced particles, enhanced production of
strangeness and of baryon/antibaryons and enhanced correlations
between particles produced in the forward and backward rapidity
hemispheres. Nevertheless, these effects are mainly noticeable at
central rapidities, the very forward and very backward regions being
only slightly affected \cite{psm2}. Results of the program PSM for
cross sections, multiplicities, longitudinal and transverse momentum
spectra and production of different hadron species can be found in
\cite{psm}, together with an extensive comparison to accelerator data.
In Fig. \ref{fig1} the energy distribution for secondary particles in
Iron-Nitrogen collisions at $10^{9}$ GeV laboratory energy is shown
with and without fusion.  The reduction of multiplicities for
low--energy 
secondaries due to fusion is clearly seen, while the
distribution of high-energy secondaries is very modestly affected.

An appealing possibility to extend the mechanism of string fusion,
allowing for a truly collective phase, is to consider a percolation
phase transition \cite{percolation}: When the density of strings in
the transverse space of the collision reaches a critical value, paths
of overlapping strings crossing the total available area appear and
the average cluster size increases suddenly with increasing string
density. Consequences of percolation have been studied extensively
\cite{perc2}. This phenomenon implies the existence of very strong
chromo-electric fields stretched between projectile and target color
sources, like in the model of \cite{Kapusta} which will be discussed
in Section IV.

As a last comment in this Section, let us note that fusion and
percolation of strings can be viewed as effective realizations of a
collective mechanism of interaction between the exchanged Pomerons, as
those introduced in \cite{nexus,dpmjet}. This mechanism also leads to
a reduction of multiplicities.

\section{Air showers and string fusion}
 
We have implemented the PSM with/without string fusion (program
PSM-1.0 \cite{psm}) in the Air shower simulation program AIRES
\cite{Aires}.  In our implementation, the possibility of string fusion
is considered only for the first nucleus-nucleus collision. This is a
valid approximation since the energy of the secondary collisions is
greatly reduced compared to the initial energy and also because these
secondary collisions take place mainly between individual nucleons and
mesons and air nuclei; thus, the density of strings is reduced in the
secondary collisions making the probability of fusion negligible.  
The AIRES hadronic model only takes into
account kaons, nucleons, and pions. All the other hadronic particles
produced in the PSM (or QGSJET \cite{QGSjet}) are forced to decay
before being passed to the AIRES subroutines.

We have run our code for iron initiated showers with energies from $10
^{14}$ to $ 10^{20}$ eV and with a relative thinning energy (see \cite{Aires})
of $10^{-6}$. 500 showers for each energy have been generated both
with and without string fusion.

\begin{figure}  
\ybox{0.4}{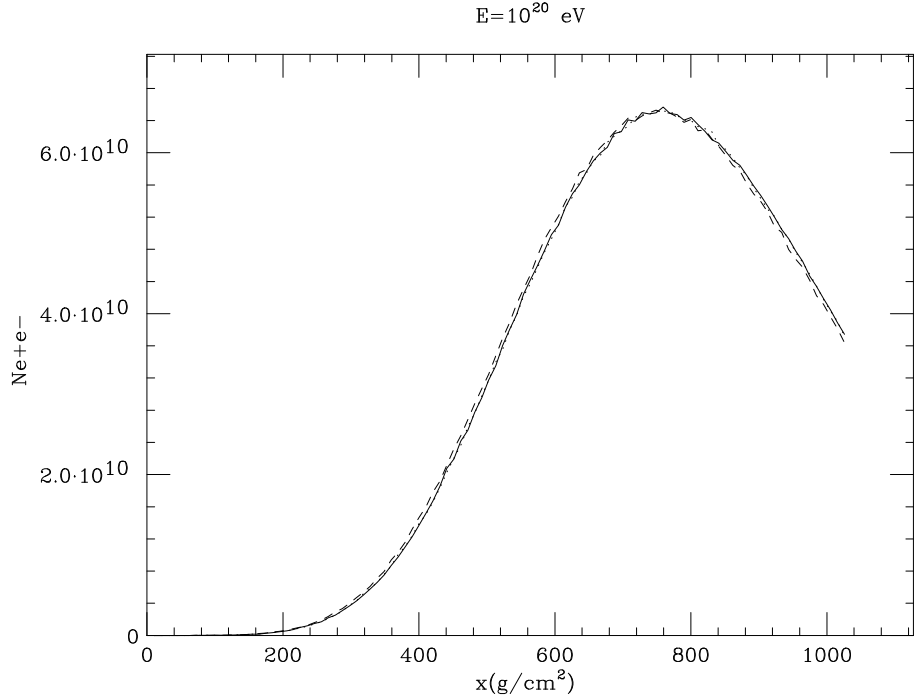}  
\caption{Average $e^+ e^-$ longitudinal development for iron 
 showers of $10^{20}$
  eV in the case with fusion (dashed line) and without fusion
  (continuous line); also results from QGSJET are shown (dotted line).}
\label{fig:edevel}
\end{figure}

\begin{figure}  
\ybox{0.4}{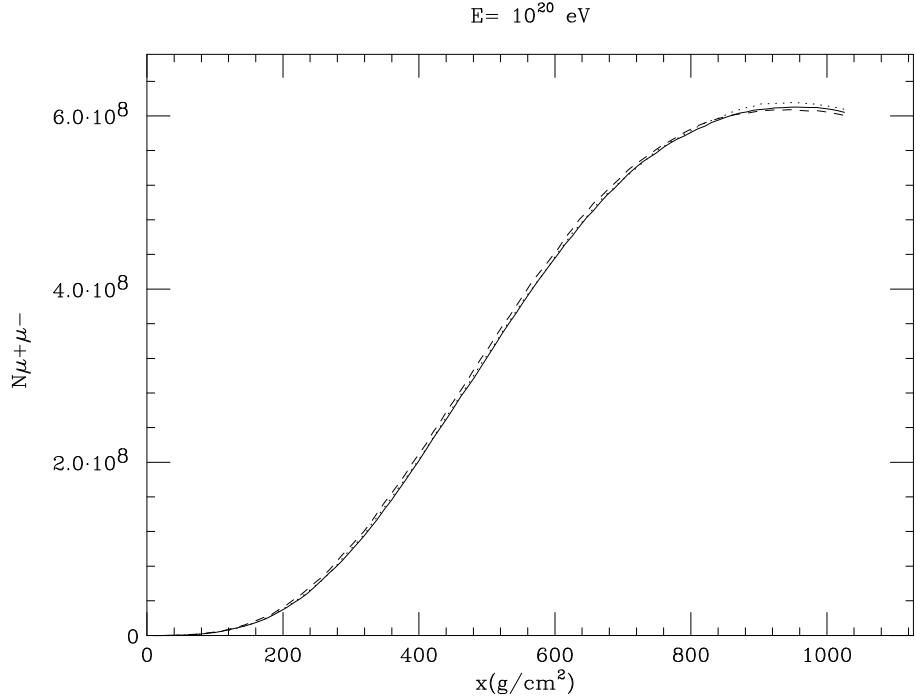}  
\caption{Average $\mu^+ \mu^-$ longitudinal development for 
 iron showers of $10^{20}$ eV, with the same line convention as in Fig.
\protect{\ref{fig:edevel}}.}
\label{fig:mudevel}
\end{figure}

Our results can be seen in Fig. \ref{fig:edevel}-\ref{fig:mudevel}.
In Fig. \ref{fig:edevel} we show the $X_{\rm max}$ distribution
for showers of $10^{20}$ eV, with and without fusion. The difference
in the average value of the $X_{\rm max}$ is less than 0.6\%,
compatible with a null result, despite the fact that at these energies
the difference in multiplicity of the first collision is about a
40\%. The average shower development are indistinguishable, as can be
seen in Fig. \ref{fig:edevel} for the longitudinal development of
electrons and in Fig. \ref{fig:mudevel} for that of the muons.

These results seem to contradict previous estimations of the effect 
\cite{Pajares,Sousa} and also the elongation rate theorem
\cite{Linsley}. However both the elongation rate theorem and the
calculations of references \cite{Pajares,Sousa} assumed that
the reduction of multiplicity can be carried over the whole cascade
process. For processes like the one discussed here in which the
reduction of multiplicity is related to the density of interacting
strings, this is not expected. 
From this result, we conclude that string fusion (or any collective
effect which affects only the central rapidity region) has no or very small 
effect on the shower development. 


\section{Air showers and reduction of particle production
in the fragmentation region}

A different possibility that produces a large stopping as a
consequence of strong collective effects, was recently considered by
Mishustin and Kapusta \cite{Kapusta}.  In the center-of-mass frame a
nuclear collision can be viewed as two thin, Lorentz-contracted, slabs
of nucleons colliding and going one through each other. The collision
proceeds through the exchange of color between nucleons of the
projectile and target. After the collision the two slabs of nucleons
will move fast apart, and in the central region a strong coherent
chromo-electric field, modeled as color strings stretched between
the excited color charges in the slabs, may form. The back-reaction
of this field on the slabs will decelerate them, causing a large
stopping power. The situation resembles the movement of charged
capacitor plates on the electric field created by themselves, with a
conversion of kinetic into potential energy.  This deceleration will
continue until the field lines stretch so much that they decay into
quark-antiquark pairs and gluons produced from the vacuum via the
Schwinger mechanism, which will eventually neutralize the
chromo-electric field.

The resulting (projectile or target)
slab will have a final $\gamma$-factor given by \cite{Kapusta}
\begin{equation}
\gamma^* = \cosh y^* = \gamma_0 \left[ 1 -\frac{\tau_0}{\lambda} \left( v_0
      \sqrt{ 1+\frac{\tau_0^2}{4 \lambda^2}} -\frac{\tau_0}{2 \lambda}
      \right) \right],
\label{stopping}
\end{equation}
where $\gamma_0=\cosh{y_0}$, $v_0=\tanh{y_0}$ are respectively the
original $\gamma$-factor and velocity of the slab and $\tau_0$ is the
time from the start of the collision until the end of the deceleration
process. $\lambda$ is the characteristic deceleration length
\begin{equation}
\lambda = \frac{\epsilon \rho_0}{\epsilon_f} \; l,
\end{equation}
where $\epsilon$ is the initial energy per baryon, $\epsilon_f$ is the
energy density of the chromo-electric field, $l$ is the slab thickness
depending on the position in transverse plane (impact parameter), and
$\rho_0$ is the nuclear density ($\rho_0l$ being the nuclear profile
function normalized to the mass number $A$).  Following Reference
\cite{Kapusta} we parameterize the energy density on the
chromo-electric field by
\begin{equation}
\epsilon_f = \epsilon_0 \left(\frac{s}{s_0}\right)^{\alpha/2} 
\left(\frac{N_p N_t}{N_0^2}\right)^\beta,
\end{equation}
where $\alpha \simeq 0.3$ is given by the low-$x$ structure function
behavior reflecting the expected increase in the number of produced
partons, and $\epsilon_0$ sets the energy scale.  $s$ is the total
center-of-mass energy squared and $s_0=1$ GeV$^2$ is a reference
energy squared. $N_{p,t}$ is the participant nucleon density of
projectile and target, depending on the impact parameter of the
collision, and $N_0 \simeq 0.4$ fm$^{-2}$.  The amount of stopping
power depends on the initial energy and also on the impact
parameter. It is expected that $\epsilon_f$ will be proportional to
the number of binary parton collisions, which here is parameterized
through the dependence on the product $N_pN_t$. For uncorrelated
collisions $\beta \simeq 1$, while in the case of strong correlations
(like those assumed in percolation of strings \cite{perc2}) one
expects $\beta \simeq 0.5\div 0.7$.

In this model the results depend on the product $\epsilon_0\tau_0$. With these
parameters chosen such that $\langle \epsilon_f\rangle \tau_0\simeq 6$
GeV/fm$^2$ for central Au Au collisions at center-of-mass energy 65 GeV
per nucleon, one gets a final center-of-mass energy of 3.5 GeV per
baryon \cite{Kapusta}. 
This can explain the strong stopping power already observed at
RHIC \cite{RHIC}.

The participant nucleon density $N_p,N_t$ depends on the impact
parameter of the collision. The overlapping area of the collision is
calculated integrating the profile function of the colliding nucleus. 
We parameterize the nuclear density using a 3 parameter Fermi
distribution taken from \cite{Jager}.

\begin{figure}  
\ybox{0.4}{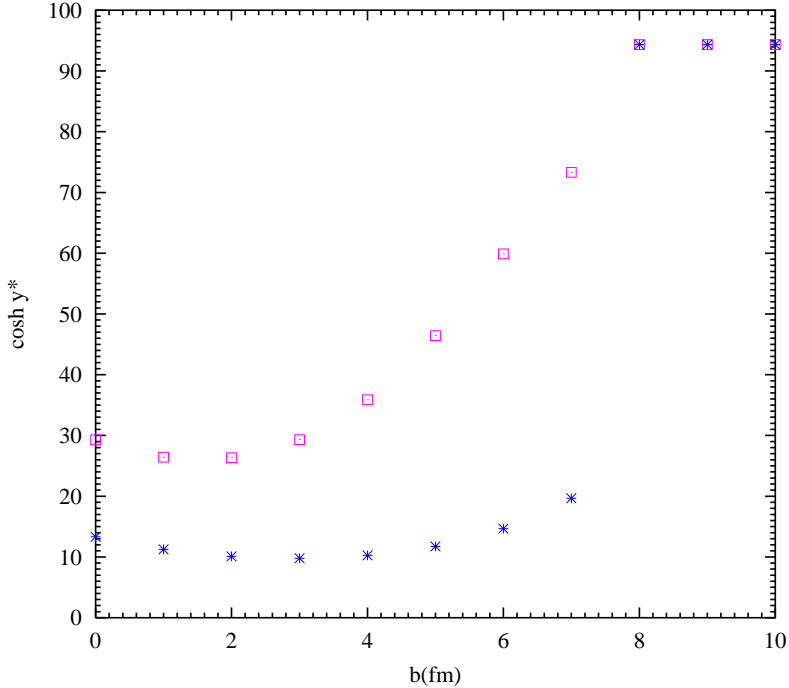}  
\caption{Forward secondary nucleon $\gamma$-factor ($\cosh y^*$) 
as a function of the impact parameter for Fe-Air collisions at
center-of-mass energy $\sqrt s = 200 $ GeV per nucleon 
for the case $\beta=1$ (squares) and $\beta=0.5$ (asterisks).}
\label{fig:coshya}
\end{figure}

The reduction in rapidity can be seen in Fig. \ref{fig:coshya} where we show
the $\gamma$-factor of the secondary nucleons ($\cosh y^*$) as a
function of impact parameter for Iron-Nitrogen collisions at
energy $\sqrt s = 200$ AGeV. As can be seen in the figure, for
small impact parameters the reduction factor is large. For central
collisions we obtain that nucleons loose more than 60 \% of their
energy at $E_{\rm lab} = 10^{17}$ eV and more than 80 \% for 10$^{19}$
eV. 

\begin{figure}  
\ybox{0.4}{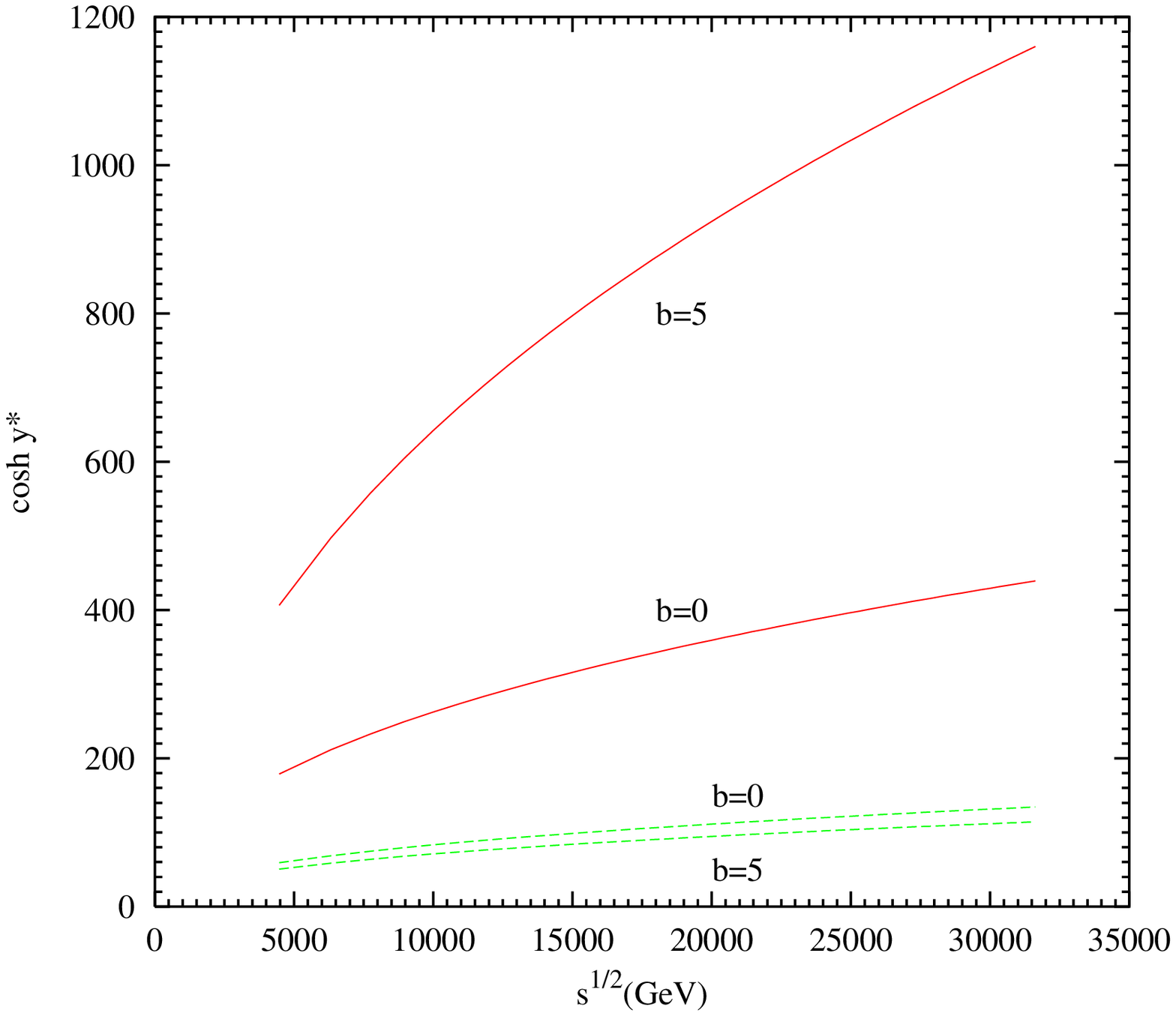}  
\caption{Forward secondary nucleon $\gamma$-factor ($\cosh y^*$) 
as a function of the center-of-mass collision energy per nucleon 
for Fe-Air collisions at fixed impact parameter (in fermi as marked) 
and $\beta=1$ (upper curves) and $\beta=0.5$ (lower curves).}
\label{fig:eloss}
\end{figure}

The effect can be further seen in Fig. \ref{fig:eloss}
where we show the same $\gamma$-factor
as a function of the energy for fixed impact parameter. This energy
loss is compatible with the measurements for Au Au collisions made at
BRAHMS \cite{RHIC}.

\begin{figure}  
\ybox{0.4}{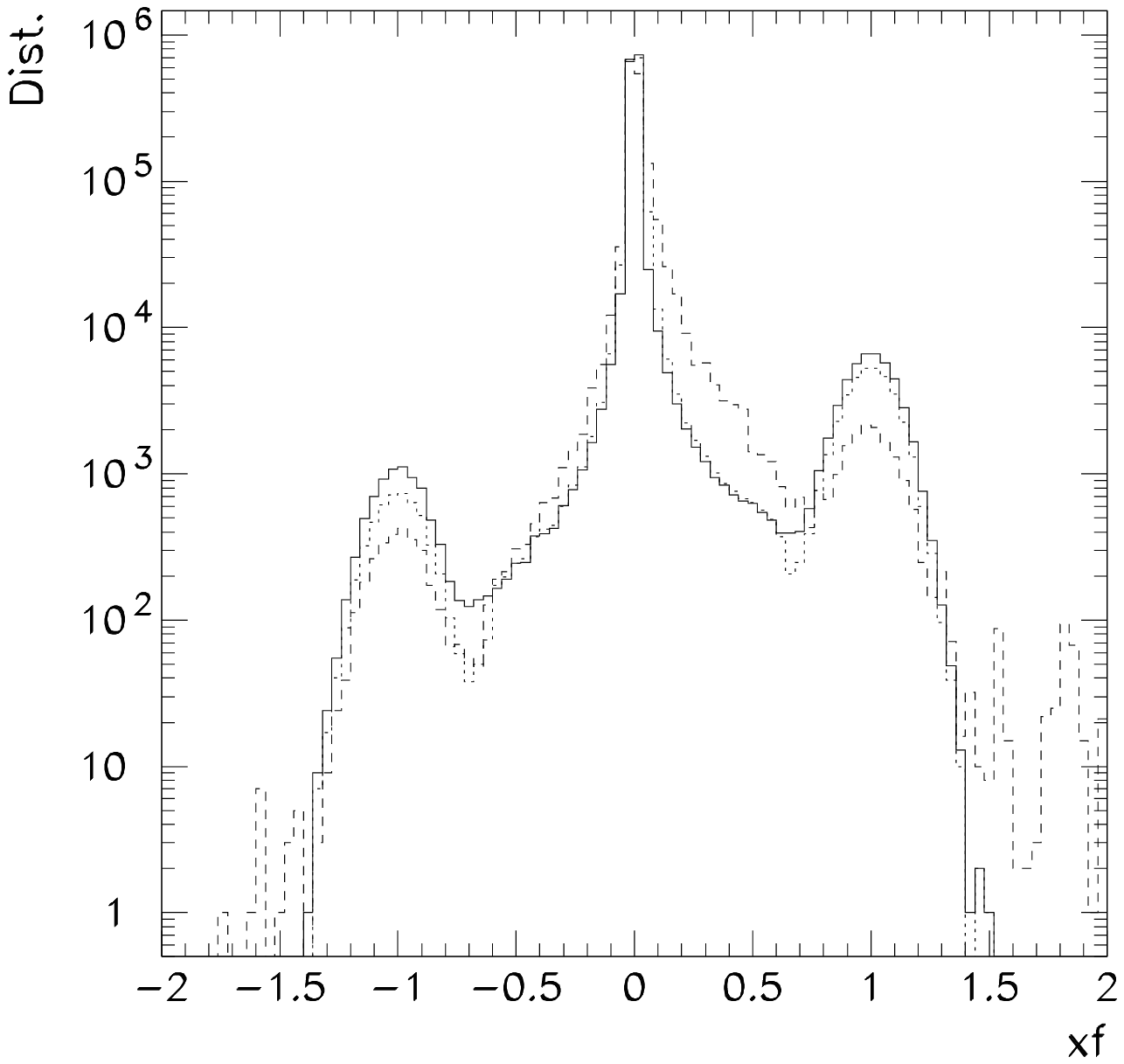}  
\caption{Center-of-mass
Feynman $x$ distribution of secondary particles in Fe-Air
  collisions at $10^{18}$ eV for the standard QGSJET model (continuous
  line), K1 (dashed line) and K2 (dotted line).}
\label{fig:xfkapusta}
\end{figure}

In Fig. \ref{fig:xfkapusta} we show the center-of-mass
$x_F$ distribution for Fe-Air
collisions at $10^{18}$ eV in the case of increase and with no
increase of the stopping power. The forward and backward peaks are
clearly reduced. 

We have implemented this increasing of the stopping power in AIRES
\cite{Aires}. In a given nucleus-Air interactions, the impact
parameter is chosen randomly and the stopping power is calculated
according to Eq. \ref{stopping}. The collision is generated with a
conventional Monte Carlo generator (QGSJET and Sibyll are implemented
in AIRES). The energy (and momentum) of the leading particles obtained
in the collision is reduced to match the stopping power predicted and
this energy, taken from the leading particles, is redistributed between
the non-leading particles.  Two different version have been essayed,
which probably bracket the possible range of these models:
in version K1 we have made the reduction for all nucleons; in version
K2, the reduction has been performed on only the participant nucleons
of each nuclei.  For the rest of the showering process the collisions
are calculated normally, so that only the first nucleus-nucleus
collision is modified.  For very high energy, $E> 10^{17}$ eV, and
small impact parameters, $b < 6$ fm, the inelasticity increase due to
this mechanism is so high that the first collision is
indistinguishable from a proton-air collision in terms of
inelasticity.  This has the effect of modifying slightly the average
$X_{\rm max}$ of the shower and the total number of muons.

\begin{figure}  
\ybox{0.4}{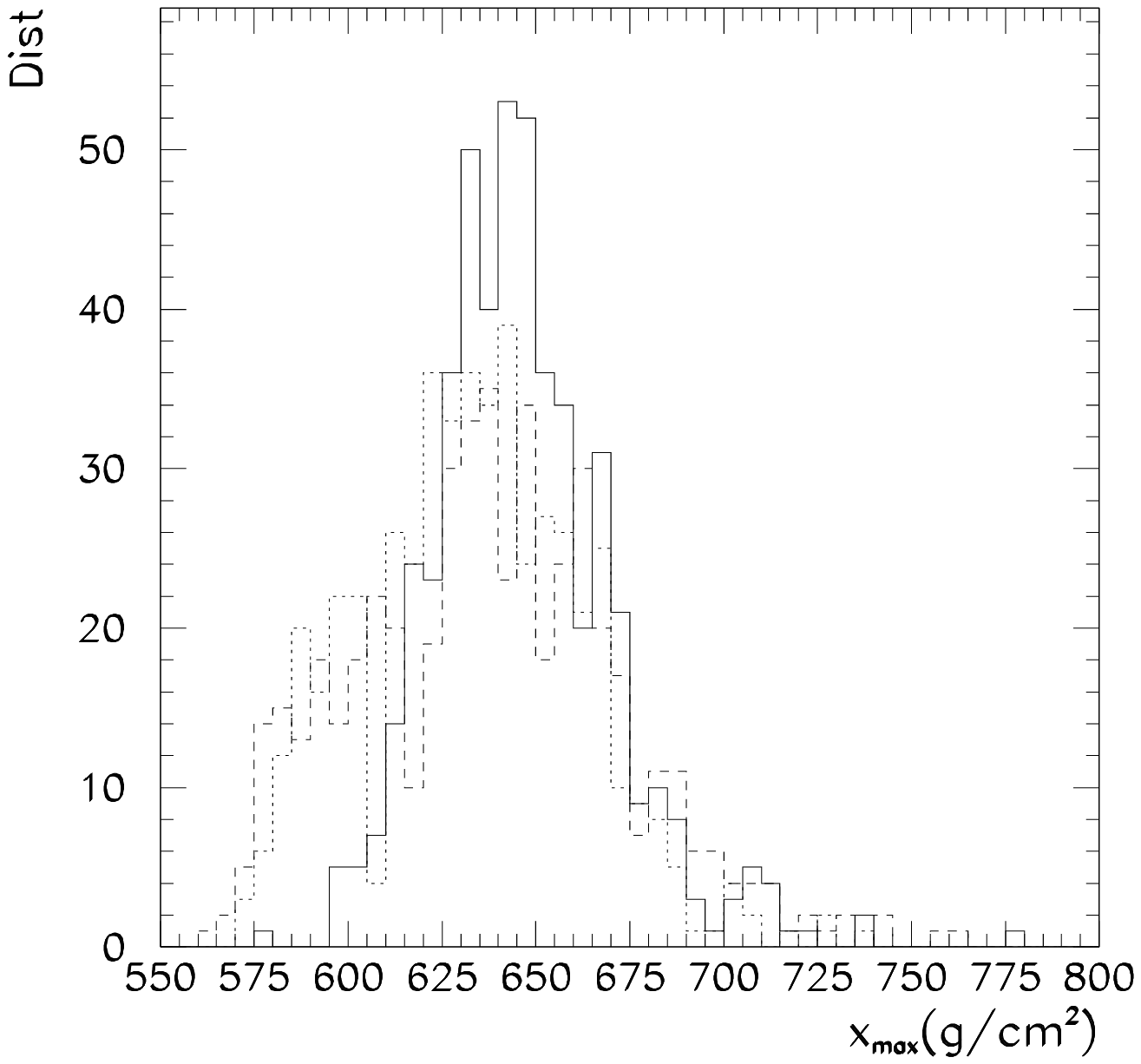}  
\caption{Shower maximum distribution for Iron showers at $E=10^{18}$eV 
for standard QGSJET (continuous line), K1 (dashed line) and K2 (dotted line).}
\label{fig:xmaxdist_k}
\end{figure}

\begin{figure}[ht]
\ybox{0.4}{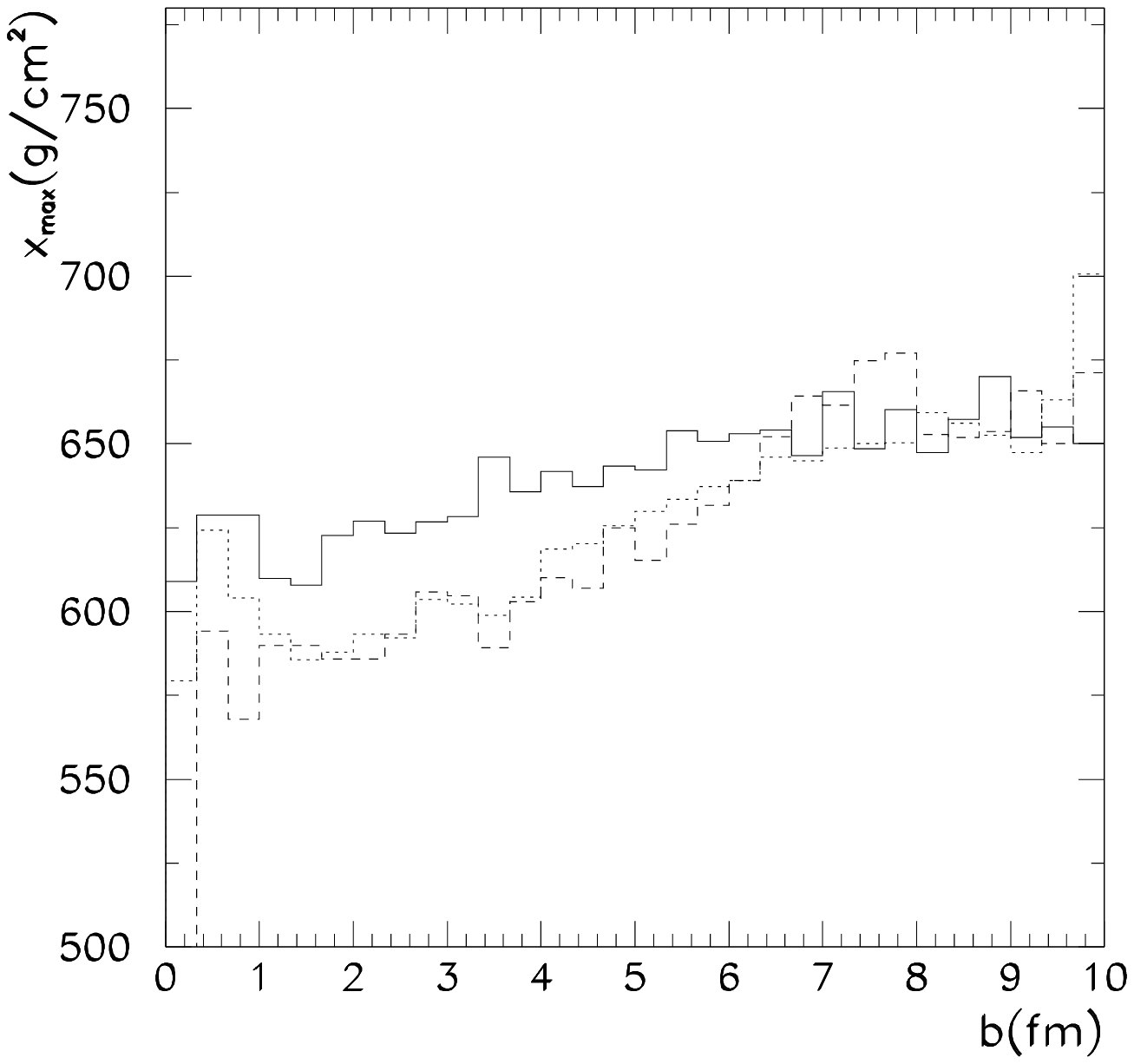}
\caption{Correlation between first--interaction 
impact parameter $b$ and shower maximum  $X_{\rm max}$ for $10^{18}$
eV iron initiated showers using the QGSJET model (continuous line) and
the stopping power enhancement mechanism in the K1 case (dashed line)
and  K2 (dotted line).}
\label{fig:xmax_k}
\end{figure}

In Fig. \ref{fig:xmaxdist_k}
we show the distribution of shower maximum for 10$^{18}$ eV iron
showers.  
Although the average value of $X_{\rm max}$ does not change much,
($\sim 13$ g/cm$^2$ for the K1 case and $\sim 8$ g/cm$^2$ for the K2 case) 
the distribution broadens, almost a factor 1.5 for the K1 case.
This is mainly due to the dependence of the strength of the elasticity
suppression on the impact parameter. 
To illustrate this point we show 
in Fig. \ref{fig:xmax_k}
the correlation between the average $X_{\rm max}$ and the impact parameter
of the collision. Collisions with a small first--interaction impact
parameter produce showers with a lower $X_{\rm max}$. Due to the high
degree of inelasticity of these collisions, these showers develop
faster. The effect is further enhanced in the model of increased
stopping power in which 
the created fields are large and the stopping
power is also large.
At large impact
parameter the effect of this mechanism is negligible and the result with and
without the modified stopping power is the same.
In this way the correlation between the first--interaction
impact parameter and $X_{\rm max}$ is increased as can be seen in the figure. 

\begin{figure}[ht]
\ybox{0.4}{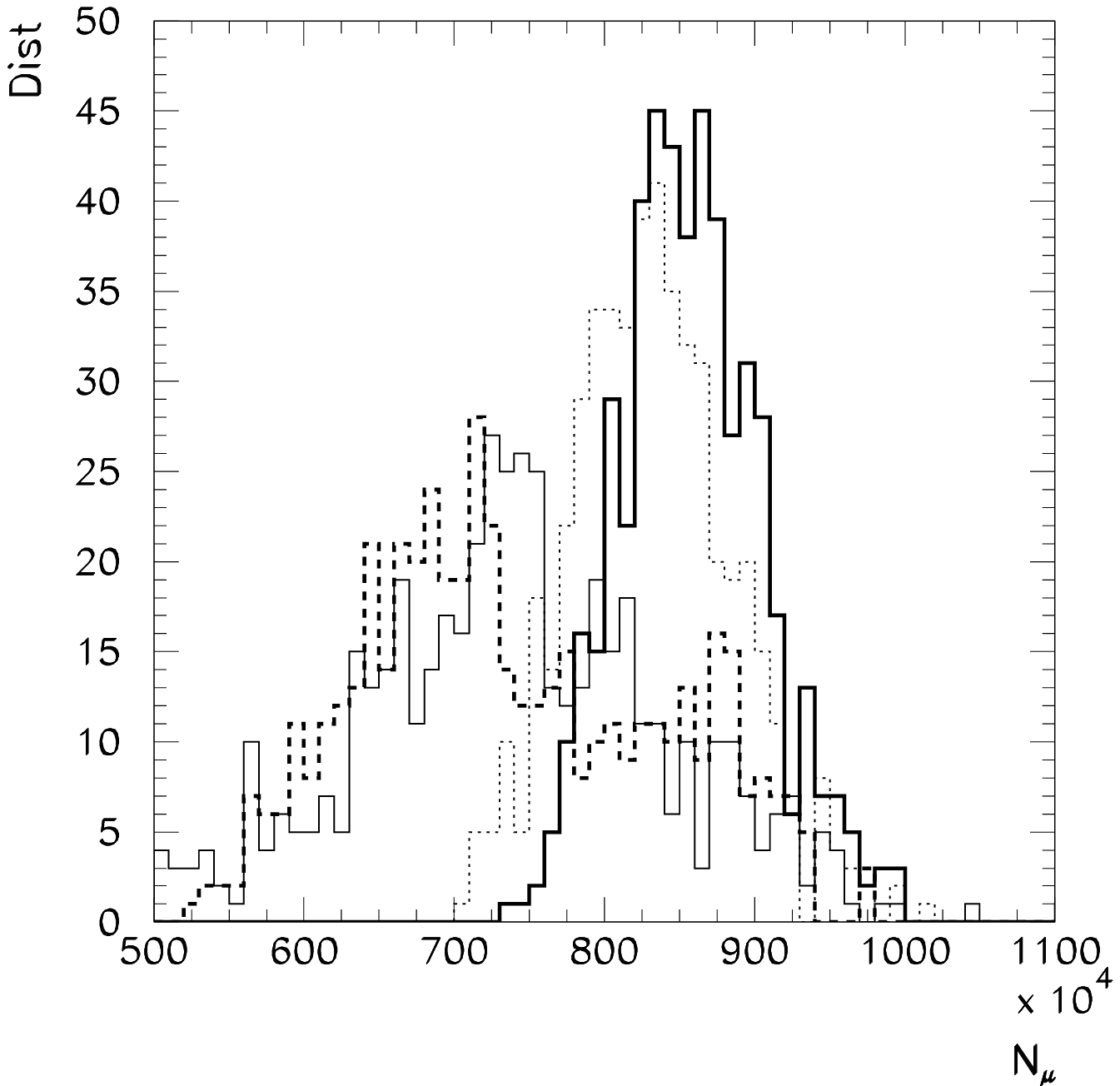}
\caption{Muon number distribution at ground for $10^{18}$
eV iron initiated showers for standard QGSJET (thick continuous line), 
K1 (thick dashed line) and K2 (dotted line). Also shown the muon number
distribution for proton initiated showers at the same energy (thin
continuous line).}
\label{fig2:satu}
\end{figure}

Finally let us turn to muon distributions. In
Fig. \ref{fig2:satu} we show the muon number distribution
at ground. Here a dramatic effect is observed for showers suffering
from this effect or not, the number of muons resembling that of a
proton shower or a regular iron shower respectively.
Both the average value and the RMS of the distribution change
appreciably. In the K1 scenario the average number of muons is $7.3
\times 10^{6}$ compared to $8.6 \times 10^{6}$ for QGSJET
iron-initiated showers. As a comparison, the average value of the
number of muons at ground for showers initiated by protons is $7.4
\times 10^{6}$. In the K2 scenario the change in the distribution of
muons at ground is small. 

\begin{figure}   
\ybox{0.4}{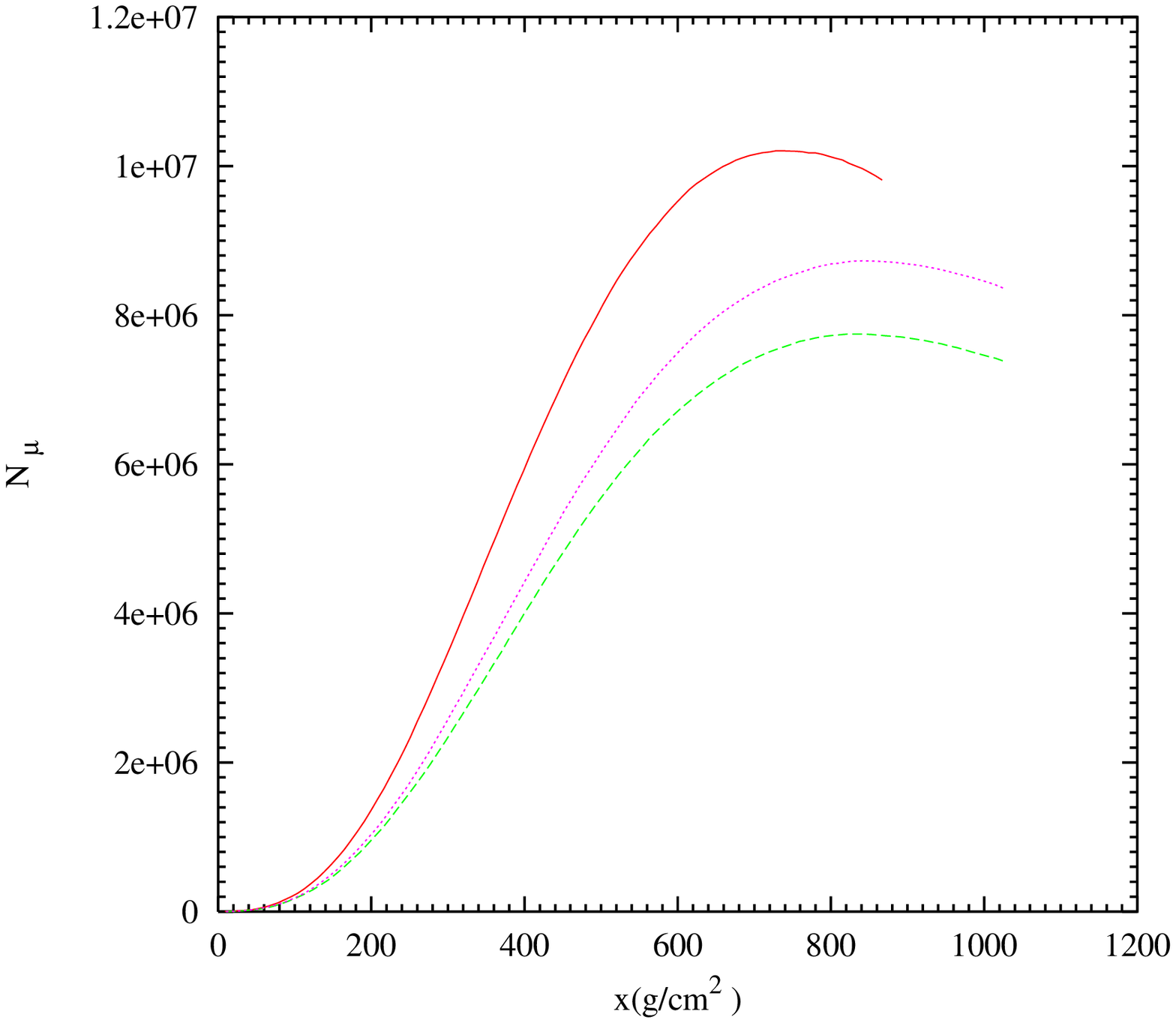}  
\caption{Average $\mu^+ \mu^-$ longitudinal development for iron
  showers of $10^{18}$ eV in standard QGSJET (continuous line), K1
  (dashed line) and K2 (dotted line).}
\label{fig5}
\end{figure}

In Figure \ref{fig5} we show the average longitudinal development for
the muon component in the standard QGSJET and K1 and K2 cases. The
muon component for the K1 and K2 cases is retarded with respect to the
standard QGSJET. 

So we observe two opposite effects of the increased stopping
scenario. On the one hand, the shower development is faster than in
conventional scenarios, which makes the shower look as initiated by a
heavier primary (i.e. more Fe-like). On the other hand, the number of
muons at ground is smaller, which makes the shower look as initiated
by a lighter projectile (i.e. more proton-like). 
This distorts the $N_\mu$--$X_{\rm max}$ relation used conventionally to
discriminate composition in an (possibly) energy dependent way.  Also
it introduces a source of systematics in the energy determination
based on the muon number such as inclined showers. The effect may be
as large as the proton--Iron difference in the number of muons. 

\section{Conclusions}

The mechanism of hadronic interactions at very high energies is still
uncertain. Existing experimental data refer to the central rapidity
region, and the forward region is very weakly constrained. This
uncertainty constitutes one of the main sources of systematic error in
the study of air showers produced by UHECRs, thus limiting the energy
and composition determination crucial to clarify their origin.  In
this paper we have tried to study the consequences on shower
development of the existence of collective mechanisms in
nucleus-nucleus collisions which modify secondary particle
production. These mechanisms have been proposed
\cite{fusion,percolation,perc2,Kapusta} to explain several
experimental facts in heavy ion collisions. We have tried two extreme
scenarios which, in our view, contain the most plausible extreme
possibilities: first, string fusion \cite{fusion} which strongly
affects multiplicities but has very little effect on the forward
region; second, a different picture of strong fields \cite{Kapusta}
which strongly increases the stopping. More concretely:

\begin{itemize}
\item{We have implement the program PSM-1.0 which include collective
  effects like string fusion in a Monte Carlo that simulate air
  showers, AIRES.  In this way we have shown that string fusion has
  almost no effect on air shower development because it only changes
  the multiplicity in the central region, the development of air
  showers being determined mainly by the particles in the forward
  region.}

\item{We have shown that other collective effects like the increasing
of the stopping power in the fragmentation region can have
consequences in the shower development. The average $X_{max}$ is
moderately reduced; this faster shower development makes the shower
look as initiated by a heavier primary.  The distribution of muons at
ground is reduced, thus mimicking a lighter primary in conventional
scenarios.}

\item{Correlations between impact parameter and shower properties
may arise in these models and constitute a novel aspect to be considered.}

\end{itemize}

These effects may have important implications in the composition
measurements and separation of proton-induced from iron-induced
showers. They should also
be taken into account to estimate the systematics of
the primary energy determination based on shower profiles and number of muons
at ground.

\section{Acknowledgments}

We thank J.~\'Alvarez-Mu\~niz, R.~Engel, and E. Zas 
for suggestions and discussions.
This work is supported by Xunta de Galicia (PGIDT00PXI20615PR), by
CICYT (AEN99-0589-C02-02), and by MCYT (FPA 2001-3837). 
We also thank the ``Centro de Supercomputaci\'on de Galicia'' (CESGA) 
for computer resources.

\end{document}